\documentclass[a4paper]{jpconf}
\usepackage{graphicx}

\begin{document}
\title{
$J/\psi$ and $\psi^{\prime}$ production in proton(deuteron)-nucleus collisions: lessons from RHIC for the proton-lead LHC run}

\author{E. G. Ferreiro$^a$, F. Fleuret$^b$, J.P. Lansberg$^c$ and A.~Rakotozafindrabe$^d$}

\address{
$^a$Departamento de F{\'\i}sica de Part{\'\i}culas and IGFAE, Universidad de Santiago de Compostela, 15782 Santiago de Compostela, Spain\\ ~ \\
$^b$Laboratoire Leprince Ringuet, \'Ecole Polytechnique, CNRS/IN2P3,  91128 Palaiseau, France\\~ \\
$^c$IPNO, Universit\'e Paris-Sud, CNRS/IN2P3, F-91406, Orsay, France \\~ \\
$^d$IRFU/SPhN, CEA Saclay, 91191 Gif-sur-Yvette Cedex, France}

\ead{elena@fpaxp1.usc.es}

\begin{abstract}
We study the impact of different cold nuclear matter effects both on $J/\psi$ and $\psi^{\prime}$ production, 
among them the modification of the gluon distribution in bound nucleons, commonly known as {\it gluon shadowing}, and 
the survival probability for a bound state to escape the nucleus --{\it the nuclear absorption}.
Less conventional effects such as {\it saturation} and {\it fractional energy loss} are also discussed. 
We pay a particular attention to the recent PHENIX preliminary data on $\psi^{\prime}$ production in $d$Au collisions at $\sqrt{s}=200$ GeV, 
which show a strong suppression for central collisions,
5 times larger than the one obtained for $J/\psi$ production at the same energy.
We conclude that none of the abovementioned mechanisms can explain this experimental result.
\end{abstract}

\section{Introduction}
The properties of the production
and the absorption of quarkonium in proton(deuteron)-nucleus collisions, 
such as $J/\psi$ and $\psi^{\prime}$ suppression in $d$Au collisions at RHIC, offer the opportunity to study 
Cold Nuclear Matter (CNM) effects in the nuclear medium. 
This is of particular relevance for the understanding of QCD at high density and temperature in nucleus-nucleus collisions,
since it is in fact adventurous to claim that the charmonium family is a quark-gluon plasma (QGP) thermometer \cite{Matsui:1986dk} 
without calibrating first the CNM suppression.

We have studied the impact of different cold nuclear matter effects both on $J/\psi$ and $\psi^{\prime}$ production, namely
the survival probability for a bound state to escape the nucleus {\it --the nuclear absorption--}
and the modification of the gluon distribution in bound nucleons  {\it --gluon shadowing}. 
Alternative mechanisms such as {\it saturation} and {\it fractional energy loss} are also considered.

We have then discussed whether these CNM effects could 
explain (or not) the data on $J/\psi$ and $\psi^{\prime}$ within a coherent picture.
We have paid a special attention to the recent PHENIX preliminary 
data \cite{Sakaguchi:2012zf}
on $\psi^{\prime}$ production in $d$Au collisions at $\sqrt{s}=200$ GeV, which show a strong suppression for central collisions
in the mid-rapidity region.

\section{Propagation in the cold nuclear matter: $J/\psi$ vs $\psi^{\prime}$}
The probability for the heavy-quark pair to be broken up in 
the propagation through the nuclear medium is 
commonly known as the nuclear absorption. It is usually 
parametrised by an effective break-up cross section~$\sigma_{\mathrm{eff}}$.

In principle, the larger size of the $\psi^{\prime}$ compared to the $J/\psi$  
tells us that the $\psi^{\prime}$ should suffer more break-up than the $J/\psi$.
Nevertheless, it is important to remember that 
the relevant timescale to analyse the pair evolution is its formation time, which, following
the uncertainty principle, 
is related 
to the time needed -- in their rest frame -- to distinguish the energy levels of the 
$1S$ and $2S$ states,
$t_f=  \frac{2 M_{c\bar c}}{(M^2_{2S}-M^2_{1S})}= 2 \times 3.3$ GeV / 4 GeV$^2= 0.35$ fm 
for the $\psi$.
Moreover, $t_f$ has to be considered in the rest frame of the target nucleus, 
{\it i.e.}~the Au beam at RHIC. 
The boost factor $\gamma$ is obtained from the rapidity of the pair corrected by the Au beam rapidity 
$\gamma=\cosh(y-y_{beam}^{\rm Au})$ where $y_{beam}^{\rm Au}=-5.36$ for RHIC. 

The formation time for the different 
rapidities reached at RHIC are given in Table~\ref{tab:tf-RHIC}.
\begin{table}[htb!]
\begin{center}\setlength{\arrayrulewidth}{1pt}
\begin{tabular}{cccc|cccc}
\hline\hline
 $y$ & $\gamma(y)$ & $t_f(y)$   & \quad& & $y$  & $\gamma(y)$ & $t_f(y)$\\
\hline
-2.0 & 14.4     & 5.1 fm &   &   &  0.0 & 106      & 36.7 fm\\
-1.5 & 23.7     & 8.3 fm &   &   & +1.5 & 476      & 166 fm\\
-1.0 & 39       & 14 fm &   &   & +2.0 & 786      & 271 fm\\
\hline\hline
\end{tabular}
\caption{\small{Boost and formation-time $y$-dependence in the Au rest frame of the $\psi$ 
at $\sqrt{s_{NN}}=200$ GeV.}}
\label{tab:tf-RHIC}
\end{center}\vspace*{-0.5cm}
\end{table}
As it is shown there, in the mid and forward rapidity regions at RHIC, $t_f$ is significantly larger than  
the Au radius. This means that the $c {\bar c}$ is nearly always in a 
pre-resonant state when traversing the nuclear matter. This implies that the break-up probability should be the same 
for
the $\psi^{\prime}$ and $J/\psi$, since these 
states cannot be distinguished at the time they traverse the nucleus. 
 
\section{Gluon shadowing and antishadowing}
At high energy,
the nucleons in a nucleus {\it shadow} each other,
leading to a modification of the nuclear Parton Distribution Functions (nPDF).
This modification, in the form of a shadowing-correction factor, has to be applied to the $J/\psi$ or $\psi^{\prime}$ yield in $d$Au 
collisions. 
This shadowing factor can be expressed in terms of the ratios $R_i^A$ of the
nPDF in a nucleon of a nucleus~$A$ to the
PDF in the free nucleon. 

We shall consider three
different shadowing parametrisations: EKS98~\cite{Eskola:1998df}, EPS08~\cite{Eskola:2008ca} and
nDSg~\cite{deFlorian:2003qf} at LO, which have been implemented in 
our Glauber Monte-Carlo framework {\sf JIN}~\cite{Ferreiro:2008qj}, 
already used to describe $J/\psi$ production at RHIC~\cite{OurExtrinsicPaper,Ferreiro:2009ur,Ferreiro:2012sy}.
Our code accounts for improved kinematics corresponding to a $2\to2$ ($g+g\rightarrow c \bar{c}+g$) 
partonic process for the $\psi$ production as in the Colour-Singlet Model at LO~\cite{Brodsky:2009cf} or
the Colour-Octet Mechanism at NLO.

\begin{figure}[h!]
\begin{minipage}[t]{.33\textwidth}
\begin{center}
\includegraphics[width=1.0\textwidth]{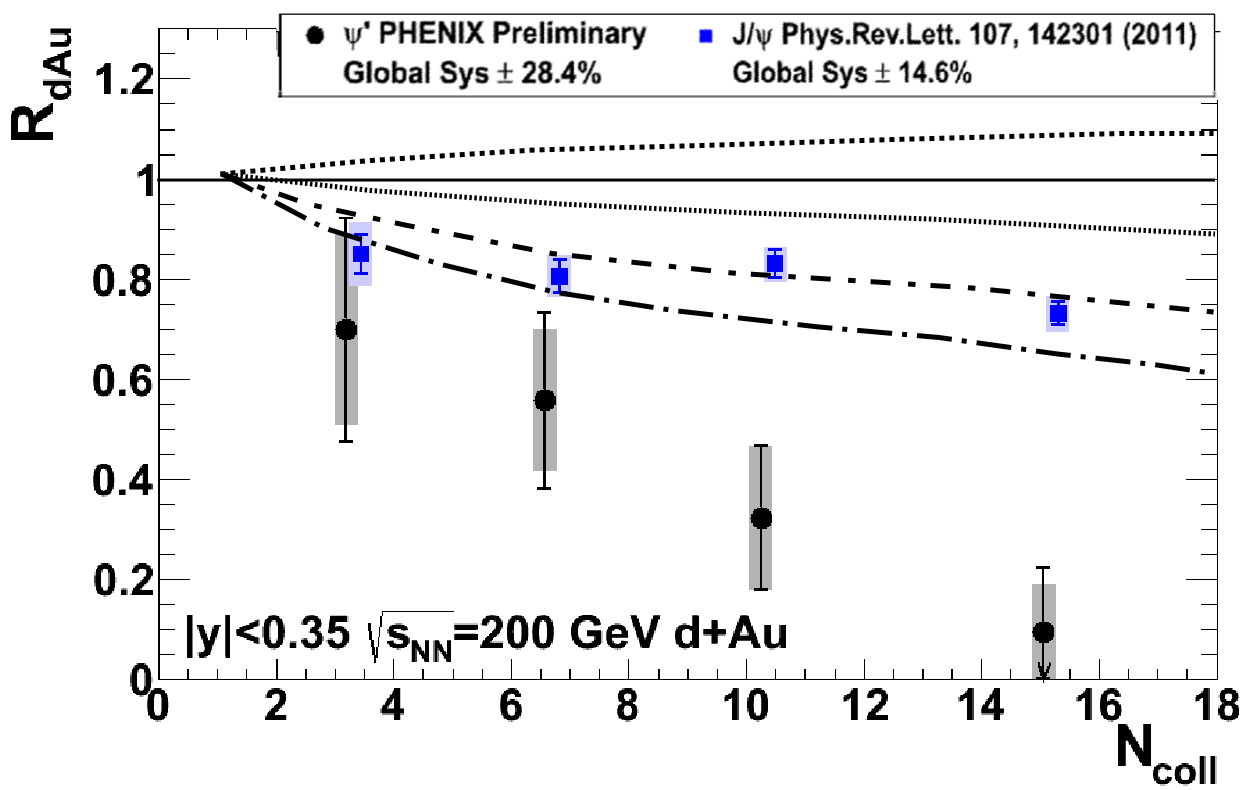}
\end{center}
\end{minipage}
\begin{minipage}[t]{.33\textwidth}
\begin{center}
\includegraphics[width=1.0\textwidth]{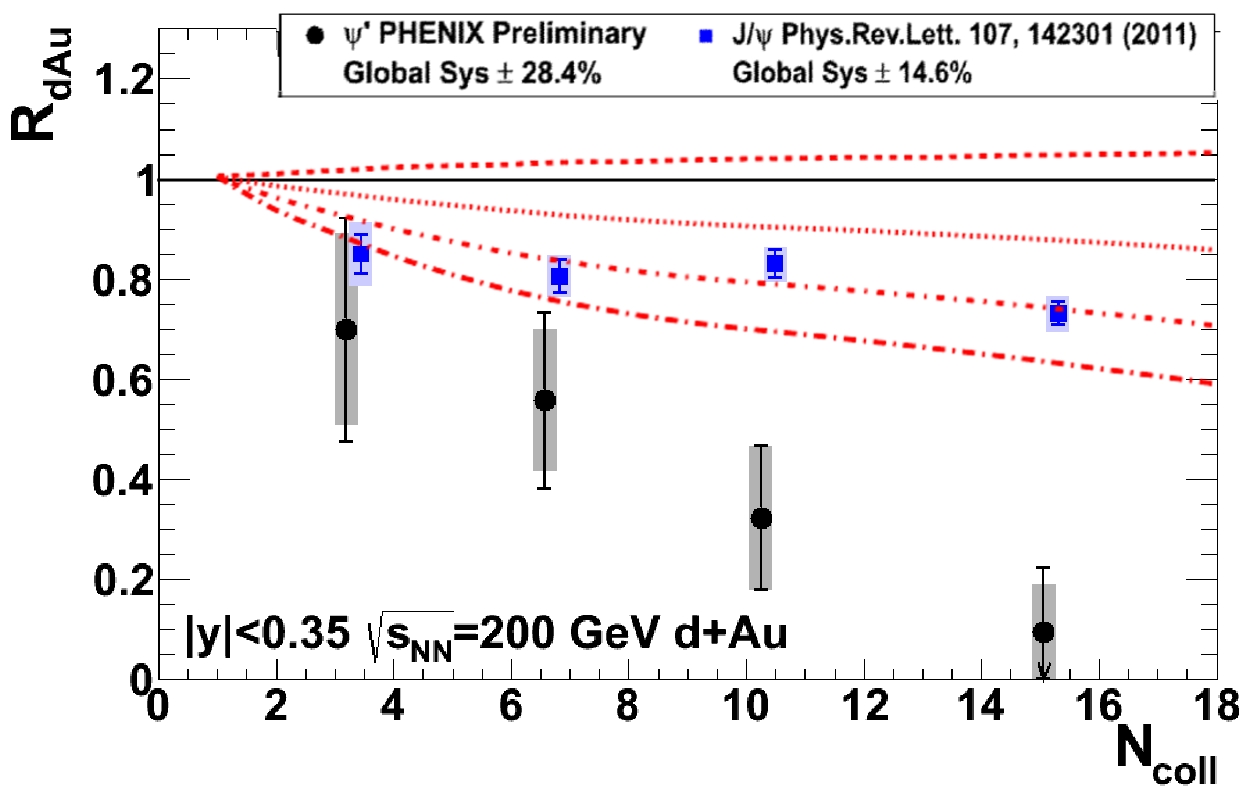}
\end{center}
\end{minipage}
\begin{minipage}[t]{.33\textwidth}
\begin{center}
\includegraphics[width=1.0\textwidth]{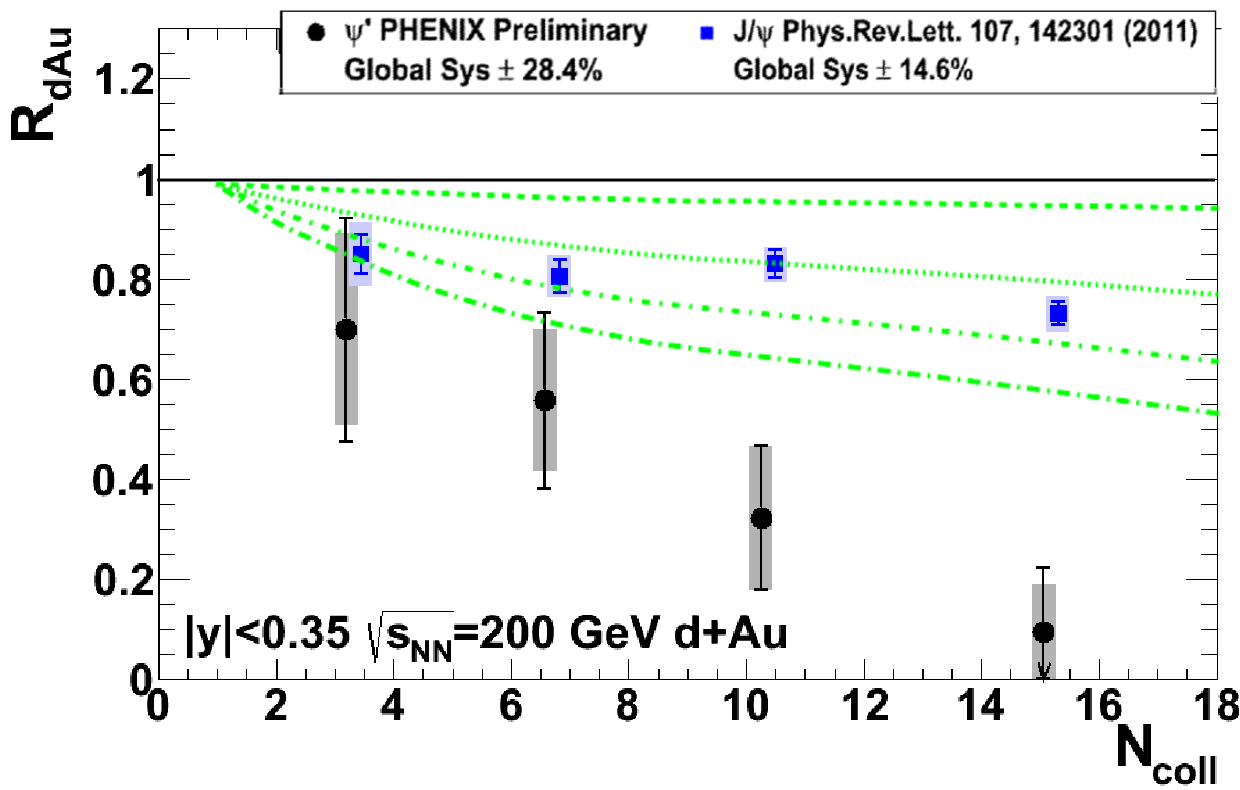}
\end{center}
\end{minipage}
\vskip -0.25cm
\caption{\small{Nuclear modification factor in $d$Au collisions, $R_{d\rm{Au}}$, 
for $J/\psi$ and $\psi^{\prime}$ as measured by PHENIX at $\sqrt{s_{NN}}=200$ GeV versus the number of 
collisions for the mid-rapidity range $|y|<0.35$ compared to our computations for four values of the nuclear absorption (from top to bottom: 
$\sigma_{\mathrm{eff}}=0,2,4,6$ mb) using EKS98 (left), EPS08 (middle) and nDSg (right).}}
\label{fig1}
\vskip -0.5cm
\hfill
\end{figure}

In Fig. \ref{fig1}, the effect of the nPDFs on $J/\psi$ is compared to PHENIX data on $J/\psi$ and $\psi^{\prime}$ production \cite{Sakaguchi:2012zf}.
We have used the three shadowing parametrisations mentioned above, together with four possibilities for the effective nuclear absortion, namely 
$\sigma_{\mathrm{eff}}=0,2,4,6$ mb.

\begin{figure}[h!]
\begin{minipage}[t]{.5\textwidth}
\begin{center}
\includegraphics[width=1.0\textwidth]{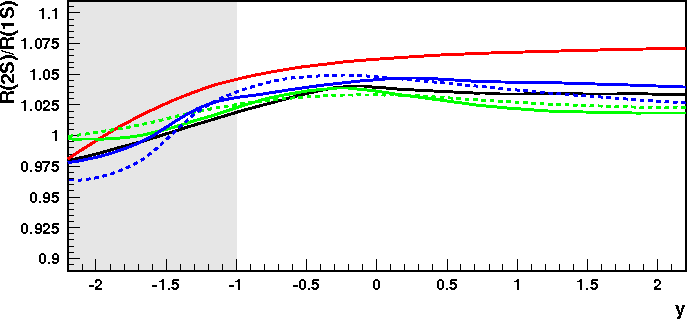}
\end{center}
\end{minipage}
\begin{minipage}[t]{.5\textwidth}
\begin{center}
\includegraphics[width=1.0\textwidth]{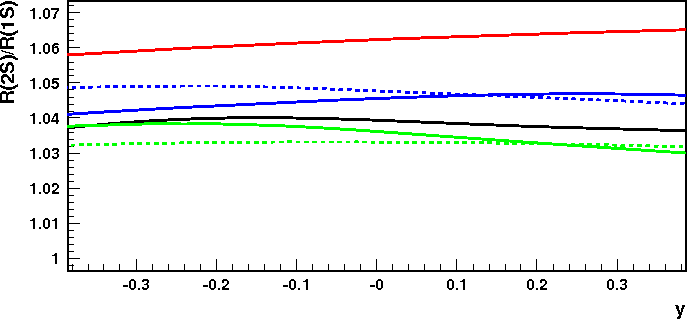}
\end{center}
\end{minipage}
\vskip -0.25cm
\caption{\small{$\psi^{\prime}$ over $J/\psi$ nuclear modification factors 
in the rapidity range $-2.2 < y < 2.2$ (left) and in the mid-rapidity region $|y|<0.35$ (right)
using EKS98 (black), EPS08 (red), nDSg (green) and EPS09 (blue) at LO (continous line) and NLO (discontinous line). The shaded
zone indicates that $t_f$ is there of the order of the Au radius. In this region, the $J/\psi$ and $\psi^{\prime}$ absorption 
may be slightly
different.}}
\label{fig2}
\vskip -0.5cm
\hfill
\end{figure}

Note that, while the choice of identical break-up cross sections for $J/\psi$ and $\psi^{\prime}$ is perfectly 
justified given the argument of large formation times invoked above, 
the common prejudice that there is a similar shadowing in both cases needs further explanations. 
It is true that the shadowing corrections, that depends on $x$ and $\mu_F$, 
should be very similar for both particle production, due to their similar masses. 
Nevertheless, the difference between their masses --of the order of $0.6$ GeV-- could induce both a shift on 
the corresponding $x$ 
at a given rapidity
and on the corresponding $\mu_F$.
We have taken 
the factorisation scale, $\mu_F$, equal
to the particle mass, 3.7 GeV for the $\psi^{\prime}$ and 3.1 GeV for the $J/\psi$.
In Fig. \ref{fig2}, our results for the ratio between the shadowing factors $R_{d\rm{Au}}^{\psi^{\prime}} \over R_{d\rm{Au}}^{J/\psi}$ for different shadowing parametrisations --EKS98, EPS08, nDSg at LO along with EPS09 at LO and NLO \cite{Eskola:2009uj}
and nDSg at NLO-- 
are shown. 
Note that the difference of the shadowing impact is at most 5 \%. Even more important, this difference will favour $\psi^{\prime}$ production in the mid-rapidity region, in contradiction to what experimental data show.

\section{Other effects: gluon saturation and fractional energy loss}
The possibility of a gluon saturation as predicted in the framework of the Color Glass Condensate (CGC) \cite{Iancu:2000hn,Ferreiro:2001qy} 
is sometimes invoked to have an impact on quarkonium production. We have already discussed \cite{Ferreiro:2011xy}
that this impact would be negligible in the case of $\Upsilon$ production at RHIC energies. 
Let us now discuss if it could affect $J/\psi$ and $\psi^{\prime}$ production.
The relevant scale below which one expects significant non-linear effects in the evolution of the gluon distributions in $p$A
is the saturation momentum. 
It can be evaluated as \cite{Albacete:2010sy}
$Q^2_{s \rm A}=A^{\frac{1}{3}}\times 0.2 \times \Big(\frac{x_0}{x}\Big)^\lambda\,  \rm{(in\, units\, of\, GeV}^2)$,
with  $\lambda\sim 0.2\div 0.3$ and $x_0=0.01$.
Typical values for RHIC kinematics
are given in Table. \ref{tab:Qs-RHIC} together with the ratio of $Q_s$ to the $\psi$ mass.
\begin{table}[htb!]
\begin{center}\setlength{\arrayrulewidth}{1pt}
\begin{tabular}{c|ccc|ccc}
\hline\hline
 $y$ 
& $Q_{s\rm Au}^{\psi^{\prime}}$(GeV) & $\frac{Q_{s\rm Au}^{\psi^{\prime}}}{m_{\psi^{\prime}}}$   
& \quad& 
& $Q_{s\rm Au}^{J/\psi}$(GeV) & $\frac{Q_{s\rm Au}^{J/\psi}}{m_{J/\psi}}$ \\
\hline
-2.2 &    $< 1$           & --  &   &&  $< 1$     & -- \\
-1.2 &   $\lsim 1$       & --  &   &&  $\lsim 1$   & -- \\
0 &   $1.0 \div 1.1$     & 0.3 &   &&   $1.0 \div 1.1$     & $0.35$ \\
1.2 &   $1.3 \div 1.4$   & $0.35 \div 0.4$  &   &&   $1.4 \div 1.5$     & $0.45 \div 0.5$ \\
2.2 &   $1.6 \div 1.9$   & $0.4 \div 0.5$  &   &&      $1.7 \div 2.0$     & $0.55 \div 0.65$ \\
\hline\hline
\end{tabular}
\caption{\small{Evaluation of the saturation scale 
in the Au nucleus, $Q_{s\rm Au}$, 
at $\sqrt{s_{NN}}=200$~GeV
in $d$Au collisions as a function of the $\psi^{\prime}$ and $J/\psi$ rapidities.}}
\label{tab:Qs-RHIC}
\end{center}\vspace*{-0.5cm}
\end{table}
Note that in the backward rapidity region $x$ is always above $x_0$ 
--the maximum momentum fraction below which parton saturation can be considered.
In the mid and forward rapidity regions, the saturation scale is always below the typical energy scale of the process, namely 
$m_{\psi^{\prime}}$ or
$m_{J/\psi}$.
In particular, in the mid-rapidity region the ratio of $Q_s$ to the $\psi$ mass is $1 \over 3$, which makes unlikely the possibility of specific saturation effects.
Only in the most forward rapidity region there is a small probability
for phenomena beyond collinear
factorisation {\it i.e.}~beyond those encoded in the nPDFs. 
Moreover, these effects --if they exist-- would
suppress the $J/\psi$ slightly more than the $\psi^{\prime}$, in contradiction to the experimental results.

Recently, the possibility of a medium-induced energy loss that scales with the energy $E$ has 
been pointed out \cite{Arleo:2010rb}.
This {\it fractional} energy loss would affect 
the heavy-quark pair produced in a coloured state provided that it has
long enough time to radiate. 
The fraction of medium-induced radiated energy is given by 
$\Delta E/E=\Delta x_1/x_1 \simeq N_c \alpha_s \sqrt{\Delta\langle p_T^2\rangle}/M_T$, 
where  $\Delta\langle p_T^2\rangle$ is the broadening of the {radiated} gluon from the proton and
$M_T$ is the transverse mass of the final-state coloured object. 
If 100\% of the $\psi$'s were produced in a coloured state at RHIC,
$\Delta x_1/x_1\sim 10 \%$. 
Assuming that 
the corresponding suppression factor is just the ratio between the PDF of the gluon 
evaluated at its original $x_1^{\prime}$, $x_1^{\prime}=x_1+\Delta x_1$ before the 
loss and the PDF at the resulting $x_1$: $R_{loss}(x_1,Q^2)=g(x_1^{\prime},Q^2)/g(x_1,Q^2)$.
Taking modern global-fit gluon {PDFs}, 
we find that this {\it {fractional} energy loss} could give an extra suppression of the order of 
10\% in the mid-rapidity region and up to 15\% in the forward region. 
Note that this suppression only applies when the $\psi$ is produced in a 
coloured state, and that its effect is similar in $J/\psi$ and $\psi^{\prime}$ production due to their
similar masses\footnote{In fact, the higher mass of the $\psi^{\prime}$ would lead to a smaller value of $\Delta x_1/x_1$, which is compensated by an increase of the mean value of $x_1$ when compared to the $J/\psi$ case. This makes the quantitative difference between the $J/\psi$ and the $\psi^{\prime}$ indistinguishable.}. Thus, this cannot explain the experimental found either. 

\vskip 0.3cm
In conclusion, we have studied different CNM effects and their impact on $J/\psi$ and $\psi^{\prime}$ production.
We have shown that, at RHIC energies and in the mid-rapidity region, the results from the four CNM  effects which we investigated, 
namely an effective nuclear absorption, the gluon shadowing, the gluon saturation and a fractional energy loss, 
impact similarly the $J/\psi$ and the $\psi^{\prime}$, in contradiction to the recent PHENIX preliminary data on 
$\psi^{\prime}$ 
production in $d$Au collisions at 200 GeV.
The abovementioned effects should be taken into account in the interpretation of the upcoming 
LHC data on $J/\psi$ and $\psi^{\prime}$ production.

\vskip 0.15cm
\noindent {\small {\bf Acknowledgments:} This work was partially supported by the Ministerio de Ciencia (Spain) \& the IN2P3 (France) (AIC-D-2011-0740).}

\end{document}